\documentclass[twocolumn,aps,prd,amsmath,amssymb]{revtex4}

\usepackage{bm}
\usepackage{amsmath}
\usepackage{amssymb}
\usepackage{latexsym}
\usepackage{amsfonts}
\usepackage{epsfig}
\usepackage{psfrag}
\usepackage{graphicx}

\usepackage{amssymb} 
\usepackage{amsmath}
\usepackage{amsfonts}
\usepackage{graphicx}

\usepackage{amssymb}
\usepackage{amscd}
\usepackage{amsmath}
\usepackage{pst-all}
\usepackage{epsf}
\usepackage{latexsym}
\usepackage{fleqn}
\usepackage{psfrag}
\usepackage{epsfig}
\usepackage{bm}

\newcommand{\ket}[1]{\left|#1\right>}
\newcommand{\bra}[1]{\left<#1\right|}

\newcommand{\ul}{\underline} 
\newcommand{\f}[1]{\mbox{\boldmath$#1$}}

\newcommand{\vau}{\mbox{\boldmath$v$}}
\newcommand{\na}{\mbox{\boldmath$\nabla$}}
\newcommand{\bea}{\begin{eqnarray}}
\newcommand{\ea}{\end{eqnarray}}
\newcommand{\eea}{\end{eqnarray}}
\newcommand{\ord}{{\mathcal O}}

\begin{document}

%\begin{frontmatter}

\title{Emergent Horizons in the Laboratory}

\author{Ralf Sch\"utzhold}

\address{Institut f\"ur Theoretische Physik,
Technische Universit\"at Dresden,
01062 Dresden, Germany}  

%email: {\sf schuetz@theory.phy.tu-dresden.de}} 

%\maketitle

\begin{abstract} 
%###
The concept of a horizon known from general relativity describes the
loss of causal connection and can be applied to non-gravitational
scenarios such as out-of-equi\-librium condensed-matter systems in the
laboratory. 
This analogy facilitates the identification and theoretical study 
(e.g., regarding the trans-Planckian problem) and possibly the
experimental verification of ``exotic'' effects known from gravity 
and cosmology, such as Hawking radiation.  
Furthermore, it yields a unified description and better understanding
of non-equilibrium phenomena in condensed matter systems and their
universal features.  
By means of several examples including general fluid flows, expanding
Bose-Einstein condensates, and dynamical quantum phase transitions, 
the concepts of event, particle, and apparent horizons will be
discussed together with the resulting quantum effects. 
\end{abstract} 

%\begin{keyword}
%analogue gravity
%\sep
%universality
%\PACS 
%04.62.+v % Quantum field theory in curved spacetime
%\sep
%04.60.-m % Quantum gravity
%\end{keyword}
%\end{frontmatter}

\maketitle

%%%%%%%%%%%%%%%%%%%%%%%%%%%%%%%%%%%%%%%%%%%%%%%%%%%%%%%%%%%%%%%%%%%%%%%%%%%%%%%
%%%%%%%%%%%%%%%%%%%%%%%%%%%%%%%%%%%%%%%%%%%%%%%%%%%%%%%%%%%%%%%%%%%%%%%%%%%%%%%
\section{Motivation}
%%%%%%%%%%%%%%%%%%%%%%%%%%%%%%%%%%%%%%%%%%%%%%%%%%%%%%%%%%%%%%%%%%%%%%%%%%%%%%%
%%%%%%%%%%%%%%%%%%%%%%%%%%%%%%%%%%%%%%%%%%%%%%%%%%%%%%%%%%%%%%%%%%%%%%%%%%%%%%%

Typically, the investigation of a given condensed-matter system starts
with understanding its equilibrium configuration, e.g., the ground
state $\ket{\psi_0}$ at zero temperature. 
However, strictly speaking, this equilibrium state provides a valid
description for {\em static} situations only.
A realistic laboratory system, on the other hand, is never completely
static -- but subject to some time-dependent external conditions such
as a varying pressure or magnetic field.
As long as these external variations are sufficiently slow
%###
(in comparison with the internal response time),  
one would argue that the equilibrium state should provide a good
approximation to the real state of the system. 
One way of making this argument more precise is given by the adiabatic
theorem:
%###
For a time-dependent quantum system at zero temperature (initially)  
with a {\em discrete} spectrum 
$H(t)\ket{\psi_n(t)}=E_n(t)\ket{\psi_n(t)}$,
the actual quantum state $\ket{\psi(t)}$ as a solution of the
Schr\"odinger equation $i\partial_t\ket{\psi(t)}=H(t)\ket{\psi(t)}$
stays close to the ground state $\ket{\psi_0}$ ($\hbar=1$) 
\bea
\label{adiabatic}
\ket{\psi(t)}
\approx
\ket{\psi_0(t)}
+
\sum_{n>0}
\frac{\bra{\psi_n}\dot{H}\ket{\psi_0}}{(E_n-E_0)^2}
\,e^{i\varphi_{n}}\,\ket{\psi_n(t)}
\,,
\ea
%###
provided that the external time-dependence of the Hamiltonian $\dot H$
is slow enough compared to the internal response times of order 
$\ord(1/[E_n-E_0])$, 
i.e.,  $\bra{\psi_n}\dot{H}\ket{\psi_0}\ll(E_n-E_0)^2$ and 
$\bra{\psi_n}\ddot{H}\ket{\psi_0}\ll(E_n-E_0)^3$ etc. 
Here $e^{i\varphi_{n}}$ denotes a pure phase factor 
(including the Berry phase). 
\\
However, this adiabaticity condition 
$\bra{\psi_n}\dot{H}\ket{\psi_0}\ll(E_n-E_0)^2$ obviously works for
discrete spectra only.
For infinitely large systems (thermodynamic limit) which support the
propagation of gap-less Goldstone modes such as sound waves, for
example, this condition does not apply.
In this case, an external time-dependence typically induces
non-equilibrium phenomena at some level (e.g., large wavelengths).  
As a result, the vacuum/ground state becomes ill-defined 
(i.e., non-unique) in these non-static scenarios leading to effects
such as the creation of quasi-particles $\ket{\psi_{n>0}}$. 
Remembering the theory of quantum fields in curved space-times 
(see, e.g., \cite{Birrell}), we see that one encounters very similar
problems and phenomena there. 
As we shall see later, there are indeed deep analogies between gravity
and condensed matter, which allow us to describe their properties in
a universal way and to understand many of these non-equilibrium
phenomena in condensed matter in terms of geometrical concepts such as
horizons.  
\\
Besides the above motivation from the condensed-matter side, the
study of these analogies is also interesting from a (quantum) gravity
point of view. 
In several important phenomena of quantum fields in curved space-times
-- such as Hawking radiation or the quantum creation of the seeds for
structure formation during inflation -- the outgoing modes originate
from initial quantum fluctuations at extremely high (trans-Planckian)
energies/momenta, where the semi-classical treatment based on
classical space-times is expected to break down
(trans-Planckian problem). 
This observation poses the question of whether and how these effects
depend on the microscopic structure at these ultra-high energies. 
Since we do not really know (yet) how to answer this question within
a full theory of quantum gravity, conclusions by analogy to suitable 
condensed-matter systems -- where we do understand the microscopic
structure -- might give us some hints.  
\\
Finally, condensed-matter theory provides us with a plethora of
emergent phenomena, i.e., large-scale properties which are not part of
the microscopic structure -- but {\em emerge} at low energies.
For example, a large class of fluids with very different microscopic
structures is described by the Euler equation at long length
scales and low energies. 
Therefore, the study of condensed-matter analogues allows us to
distinguish between universal (e.g., emergent) properties and
system-specific features, which provide some information about the
underlying microscopic structure. 

%%%%%%%%%%%%%%%%%%%%%%%%%%%%%%%%%%%%%%%%%%%%%%%%%%%%%%%%%%%%%%%%%%%%%%%%%%%%%%%
%%%%%%%%%%%%%%%%%%%%%%%%%%%%%%%%%%%%%%%%%%%%%%%%%%%%%%%%%%%%%%%%%%%%%%%%%%%%%%%
\section{The underlying analogy}\label{analogy}
%%%%%%%%%%%%%%%%%%%%%%%%%%%%%%%%%%%%%%%%%%%%%%%%%%%%%%%%%%%%%%%%%%%%%%%%%%%%%%%
%%%%%%%%%%%%%%%%%%%%%%%%%%%%%%%%%%%%%%%%%%%%%%%%%%%%%%%%%%%%%%%%%%%%%%%%%%%%%%%

Motivated by the aforementioned trans-Planckian problem, Bill Unruh
suggested \cite{Dumb} studying sound waves in an irrotational and
inviscid (i.e., frictionless) flow. 
In view of the vanishing vorticity $\na\times\vau=0$, sound
perturbations can be described by a scalar potential via
$\delta\vau=\na\phi$. 
Insertion into the Euler and continuity equations yields the wave
equation for sound 
\bea
\left(\frac{\partial}{\partial t}+\na\cdot\vau_0\right)
\frac{\varrho_0}{c^2_{\rm s}}
\left(\frac{\partial}{\partial t}+\vau_0\cdot\na\right)
\phi
=
\na\cdot(\varrho_0\na\phi)
\,,
\ea
where $\varrho_0$ and $\vau_0$ denote the density and velocity of
the background flow and $c_{\rm s}$ is the speed of sound. 
Interestingly, this wave equation is identical to that of a scalar
field $\phi$ in a curved space-time 
\bea
\Box_{\rm eff}\phi
=
\frac{1}{\sqrt{-g_{\rm eff}}}\partial_\mu
\left(\sqrt{-g_{\rm eff}}\,g^{\mu\nu}_{\rm eff}\partial_\nu\phi\right)
=0
\,,
\ea
if we encode the properties of the background flow into the effective
metric $g^{\mu\nu}_{\rm eff}$ in the
Painlev{\'e}-Gullstrand-Lema{\^\i}tre form \cite{PGL} 
\bea
\label{PGL}
g^{\mu\nu}_{\rm eff}
=
\frac{1}{\varrho_0c_{\rm s}}
\left(
\begin{array}{cc}
1 & \vau_0 \\
\vau_0 & \vau_0\otimes\vau_0 - c^2_{\rm s}\f{1}
\end{array}
\right)
\,.
\ea
In summary, we arrive at the qualitative analogy (sketched below) on
the kinematic level -- which does, however, not extend to the
%###
dynamics.\footnote{Note that only the phonons ``feel'' the effective
  metric (\ref{PGL}). A detector immersed in the fluid will evolve
  according to the Minkowski metric (unless it is designed in a
  special way, see, e.g., \cite{fedichev}). Therefore, the internal
  time of the detector will be the laboratory time and hence there
  will be no direct fluid analogue of the Unruh effect
  \cite{unruh}, as this effect is a result of the relativistic time
  dilatation with a constantly changing velocity and hence Lorentz
  boost factor. Nevertheless, it is possible to design analogues for
  indirect signatures of the Unruh effect, see, e.g., 
  \cite{signatures}.}  
Of course, the background flow is governed by the Euler equation and
thus the above metric is not obtained by solving the Einstein
equations. 

\begin{center}
\begin{tabular}{ccc}
Phonons (quantised) & $\leftrightarrow$ & Quantum fields
\\
Fluid flow (classical) & $\leftrightarrow$ & Gravitational field
\\
Euler equation & $\neq$ & Einstein equations
\end{tabular}
\end{center}

%###
Having established this striking analogy between phonons in flowing
fluids and (scalar) quantum fields in curved space-times
%###
-- in spite of the in-equivalence of Euler and Einstein equations, 
one may wonder whether it is a peculiarity of sound waves or a more
general phenomenon.
As it turns out, the analogy can be extended to more general
quasi-particles under appropriate conditions:
The most general linearised low-energy effective action for scalar
Goldstone-mode quasi-particles $\phi$ can be cast into the following
form 
\bea
\label{general}
{\mathcal L}_{\rm eff}
=
\frac12(\partial_\mu\phi)(\partial_\nu\phi)G^{\mu\nu}(\ul x)
+\ord(\phi^3)
+\ord(\partial^3)
\,,
\ea
where $G^{\mu\nu}(\ul x)$ denotes some tensor depending on the
underlying condensed-matter system, see, e.g., \cite{Living,Barcelo}. 
Higher-order contributions $\ord(\phi^3)$ and $\ord(\partial^3)$ have
been omitted since we are considering a linearised and low-energy 
effective action. 
Note that additional terms like $U^\mu(\ul x)\phi\partial_\mu\phi$
would be equivalent to a potential term $V(\ul x)\phi^2$ after an
integration by parts -- which is forbidden for Goldstone (i.e.,
gap-less) modes.
(However, a potential term $V(\ul x)\phi^2$ would not spoil the 
curved space-time analogy and could be included as well -- 
as long as it is still within the low-energy realm.) 
As usual, source terms like $J(\ul x)\phi$ and 
$K^\mu(\ul x)\partial_\mu\phi$ can be absorbed by a field-redefinition
(quadratic completion). 
\\
Under these circumstances, we see that also more general
quasi-particles $\phi$ are analogous to quantum fields in curved
space-times. 
Again all relevant features of the underlying condensed-matter system
are encoded in the tensor $G^{\mu\nu}$ and hence in the effective
metric $G^{\mu\nu}\to g^{\mu\nu}_{\rm eff}\sqrt{-g_{\rm eff}}$. 
Besides phonons, examples for such quasi-particles are ripplons
(surface waves in fluids, see, e.g., \cite{ripplon}) and magnons 
(magnetisation variations, see, e.g., \cite{qpt}), provided that they
can be described by a single scalar field $\phi$.  
For multiple (non-scalar) quasi-particle modes $\phi_a$ which mix, the
tensor $G^{\mu\nu}$ acquires additional indices in general 
$G^{\mu\nu}_{ab}$ and the analogy to an effective metric breaks down
in the absence of additional symmetries. 
If these multiple quasi-particle modes are related via a symmetry,
however, it may be possible to extend the curved space-time analogy
to the non-scalar case.
One example are photons in a dielectric medium with a constant
permittivity $\varepsilon$ and a possibly space-time dependent
four-velocity $u^\mu$, see \cite{Soff} and \cite{Slow}.  
In this case, both polarisations behave exactly as in a gravitational
field described by the effective (Gordon) metric \cite{Gordon}
\bea
\label{gordon-o}
g^{\mu\nu}_{\rm eff}=g^{\mu\nu}_{\rm Minkowski}
+(\varepsilon-1)\,u^\mu\,u^\nu
\,.
\ea
%

%%%%%%%%%%%%%%%%%%%%%%%%%%%%%%%%%%%%%%%%%%%%%%%%%%%%%%%%%%%%%%%%%%%%%%%%%%%%%%%
%%%%%%%%%%%%%%%%%%%%%%%%%%%%%%%%%%%%%%%%%%%%%%%%%%%%%%%%%%%%%%%%%%%%%%%%%%%%%%%
\section{Emergent horizons}
%%%%%%%%%%%%%%%%%%%%%%%%%%%%%%%%%%%%%%%%%%%%%%%%%%%%%%%%%%%%%%%%%%%%%%%%%%%%%%%
%%%%%%%%%%%%%%%%%%%%%%%%%%%%%%%%%%%%%%%%%%%%%%%%%%%%%%%%%%%%%%%%%%%%%%%%%%%%%%%

Due to the qualitative analogy between phonons (or other
quasi-particles) and quantum fields in curved space-times, all 
kinematic aspects of the phonons are encoded in the effective metric 
$g^{\mu\nu}_{\rm eff}(\varrho_0,c_{\rm s},\vau_0)$.  
Consequently, this analogy provides a universal description (at low
energies), i.e., we can forget all the microscopic details of the
underlying fluid and apply the powerful tools and geometrical concepts
known from general relativity \cite{Wheeler} to this metric.
For example, we may introduce sound cones (in analogy to light cones)
via 
\bea
ds^2_{\rm eff}=g_{\mu\nu}^{\rm eff}\;dx^\mu\,dx^\nu=0
\,,
\ea
which determine the causal structure of the fluid -- i.e., which
points can be connected by sound waves (or other quasi-particles) 
and which cannot. 
%###
In general relativity, a break-down of causal connection (latter case) 
induced by the dynamics of the background (space-time) corresponds to
the existence of a horizon -- and this useful concept can now be
transferred to condensed-matter systems, see also \cite{Memorial}. 
%
%a dynamical loss of causal connection 
%(latter case) corresponds to the existence of a horizon, which marks a
%dynamical break-down of the causal connection, see also \cite{Memorial}. 
%
In order to be in equilibrium, every point of a condensed-matter
system must be able to exchange energy with any other point 
(e.g., for equilibrating their local temperature or pressure). 
Since such an exchange is, at low energies, mediated by
quasi-particles such as phonons, the emergence of a horizon analogue
always indicates a departure from equilibrium and entails the
amplification of quantum fluctuations (creation of quasi-particles) 
in general.  

%%%%%%%%%%%%%%%%%%%%%%%%%%%%%%%%%%%%%%%%%%%%%%%%%%%%%%%%%%%%%%%%%%%%%%%%%%%%%%%
%%%%%%%%%%%%%%%%%%%%%%%%%%%%%%%%%%%%%%%%%%%%%%%%%%%%%%%%%%%%%%%%%%%%%%%%%%%%%%%
\section{Event horizon}
%%%%%%%%%%%%%%%%%%%%%%%%%%%%%%%%%%%%%%%%%%%%%%%%%%%%%%%%%%%%%%%%%%%%%%%%%%%%%%%
%%%%%%%%%%%%%%%%%%%%%%%%%%%%%%%%%%%%%%%%%%%%%%%%%%%%%%%%%%%%%%%%%%%%%%%%%%%%%%%

\begin{figure}[ht]
%\mbox{
\epsfxsize=4.5cm\epsffile{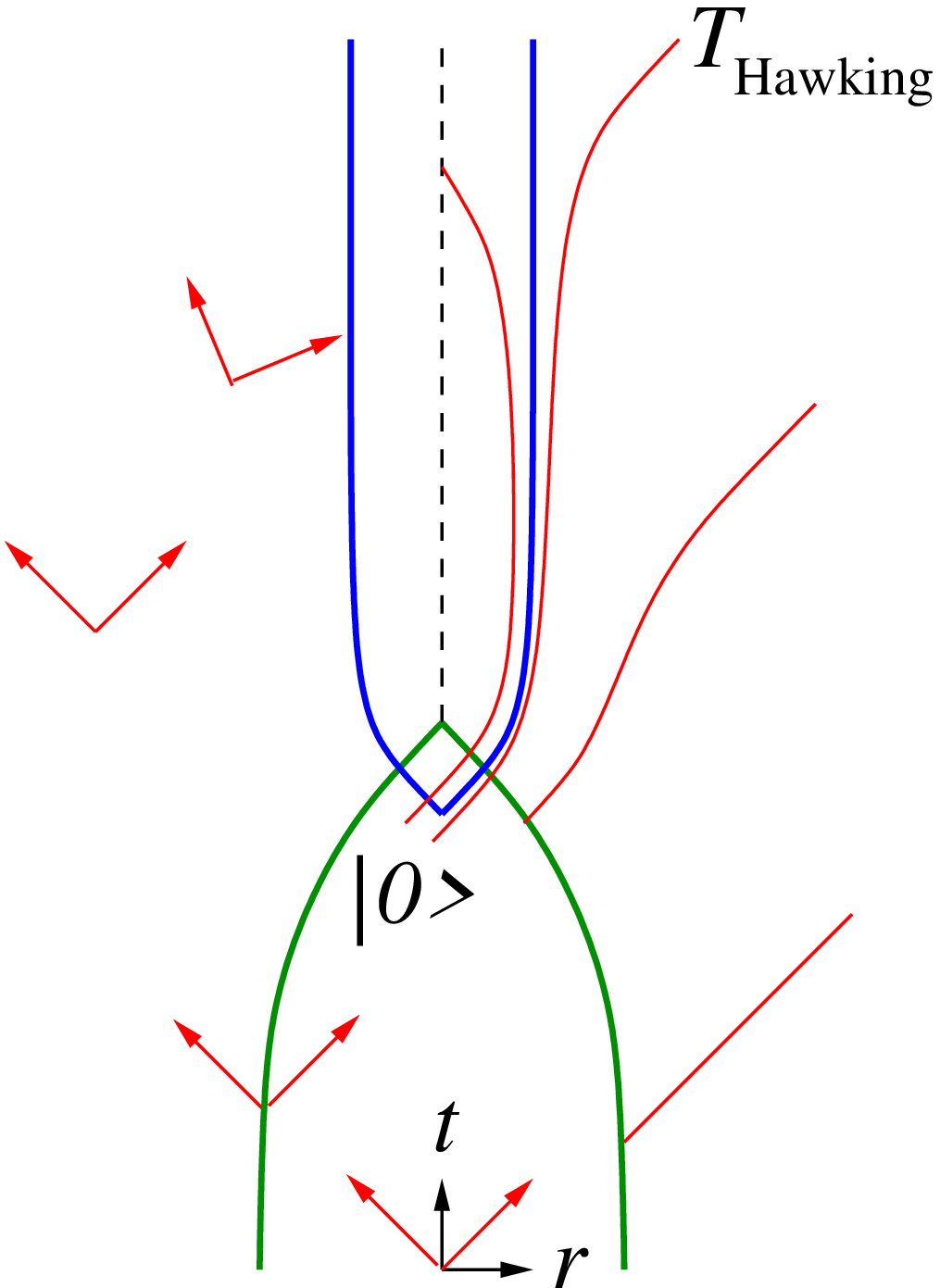}
%\hspace{.5cm}
\epsfxsize=8cm\epsffile{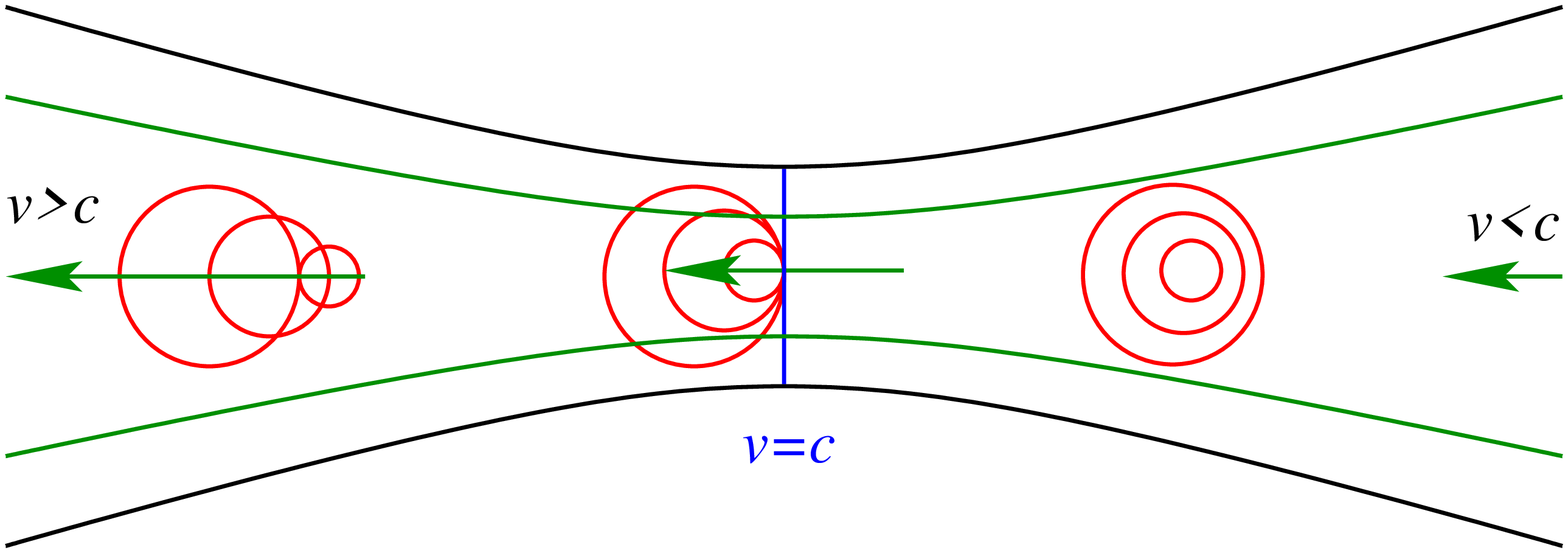}
%}
\caption{Space-time diagram of the collapse of matter to a black hole
  (top) and sketch of a de~Laval nozzle as the analogue of an event
  horizon (bottom). In the top picture, the green line indicates the
  surface of the collapsing matter and the black dashed line is the
  singularity. Light rays are denoted by red lines/arrows. The event
  horizon (blue line) lies between the last light ray which is able to
  escape to infinity and the first ray which is trapped. In the fluid
  analogue (bottom), green lines/arrows denote the fluid flow and the
  black lines are the walls of the nozzle. At its entrance, the flow
  is sub-sonic; and the flow velocity reaches the speed of sound at
  the narrowest point. Somewhat against intuition, the flow
  accelerates further (instead of slowing down) when the nozzle
  becomes wider again (as can be seen from the Bernoulli equation) and
  exits the nozzle with super-sonic speed. Sound waves (red circles)
  cannot escape from the super-sonic to the sub-sonic region and thus
  the boundary (blue line) between the two is completely analogous to
  an event horizon. 
}
\label{event}
\end{figure}

As a first example, let us consider the event horizon known from black
holes.
The event horizon is defined as the border of the region from where
nothing is able to escape to infinity, cf.~Fig.~\ref{event}.
Even though black holes are supposed to be completely black at the
classical level, the incorporation of quantum effects suggests that
they emit thermal radiation (black hole evaporation) as determined by
the Hawking temperature \cite{Hawking}
\bea
T_{\rm Hawking}
=
\frac{1}{8\pi M}
\frac{\hbar\,c^3}{G_{\rm N}k_{\rm B}}
\,,
\ea
where $M$ is the (ADM) mass of the black hole. 
The Hawking effect provides a striking confirmation of Bekenstein's
concept of black hole thermodynamics \cite{Bekenstein}.
Extending the second law of thermodynamics to black holes then
requires assigning a non-vanishing entropy to the black hole 
$S_{\rm BH}=k_{\rm B} A/(4L_{\rm Planck}^2)$, which is given by the 
horizon area $A$ in units of the Planck length $L_{\rm Planck}$ 
squared. 
Understanding the meaning and origin of this entropy is one of the
important questions a theory of quantum gravity should answer. 

However, there is a potential flaw in this picture:
Tracing the outgoing particles of the Hawking radiation back in time
(and thereby undoing the immense gravitational red-shift near the 
horizon), one finds that they originate from modes with ultra-short
wavelengths, which are far below the Planck length $L_{\rm Planck}$ for
realistic parameters. 
However, at such ultra-short length scales, one would not trust the
semi-classical treatment of quantum fields propagating in classical
space-times anymore.
Instead, effects of quantum gravity are expected to become important
at such scales.
Therefore, the derivation of Hawking radiation relies on the
extrapolation of a theory into a region where it is expected to break  
down. 
Ergo, one arrives at the question of whether the Hawking effect is
just an artifact of this unphysical over-extrapolation or whether it 
survives even after taking into account the proper microscopic
structure (trans-Planckian problem).  

%%%%%%%%%%%%%%%%%%%%%%%%%%%%%%%%%%%%%%%%%%%%%%%%%%%%%%%%%%%%%%%%%%%%%%%%%%%%%%%
%%%%%%%%%%%%%%%%%%%%%%%%%%%%%%%%%%%%%%%%%%%%%%%%%%%%%%%%%%%%%%%%%%%%%%%%%%%%%%%
\section{Impact of dispersion relation}
%%%%%%%%%%%%%%%%%%%%%%%%%%%%%%%%%%%%%%%%%%%%%%%%%%%%%%%%%%%%%%%%%%%%%%%%%%%%%%%
%%%%%%%%%%%%%%%%%%%%%%%%%%%%%%%%%%%%%%%%%%%%%%%%%%%%%%%%%%%%%%%%%%%%%%%%%%%%%%%

\begin{figure}[ht]
\begin{center}
\mbox{\epsfxsize=8cm\epsffile{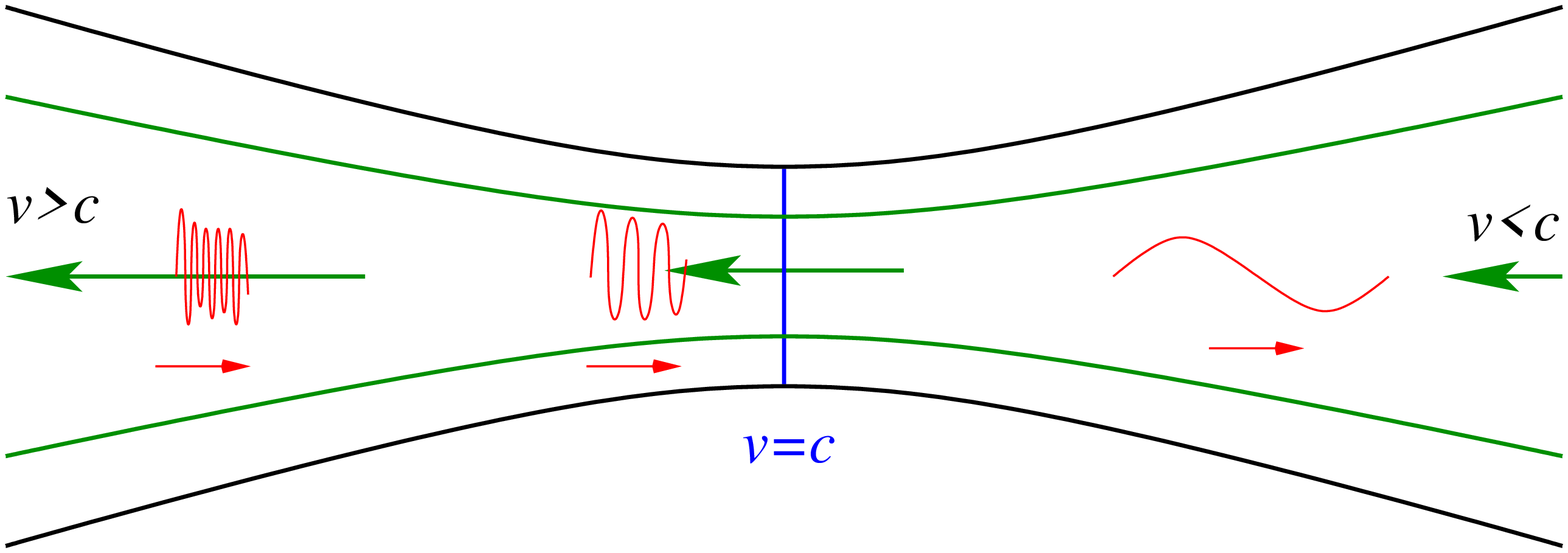}}
\end{center}
\caption{Sketch of the origin of Hawking radiation (red curves) in a 
  de~Laval nozzle.} 
\label{laval-BEC}
\end{figure}

Since we do not know (yet) the microscopic structure at the Planck
scale (and beyond), it is very hard to answer this important question
for real gravity.
Fortunately, the analogy sketched in Fig.~\ref{event} yields a toy
model which captures most of the relevant features and is still simple
enough that we are able to do the calculations in certain cases. 
In view of the qualitative analogy in Sec.~\ref{analogy}, one would
also expect the de~Laval nozzle to emit Hawking radiation:
Even though the in-flowing fluid is at zero temperature, there should
be a thermal flux of phonons (a faint hissing noise) escaping to the
right-hand side. 
Repeating the same derivation for the metric (\ref{PGL}) instead of
the Schwarzschild geometry, the (analogue) Hawking temperature is
determined by the velocity gradient at the horizon 
(where $v_0=c_{\rm s}$), see \cite{Dumb}
\bea
T_{\rm Hawking}
=
\frac{\hbar}{2\pi\,k_{\rm B}}\,
\left|\frac{\partial}{\partial r}\left(v_0-c_{\rm s}\right)\right|
\,.
\ea
Depending on the experimental realisation, this temperature could
range from a few nano-Kelvin (for Bose-Einstein condensates, see,
e.g., \cite{Garay,Raizen}) up to fractions of a Kelvin (e.g., for
wave-guides, cf.~\cite{wave-guide}).  
These values are much larger than the Hawking temperatures of real
solar-size black holes, but still very hard to measure \cite{pessi}.  
Besides a possible experimental realisation, we may also use the fluid
analogues as toy models for quantum gravity in order to address the
trans-Planckian problem. 
Again, the outgoing long-wavelength phonons originate from
short-wavelength modes near the horizon -- for which the fluid
description breaks down. 
Fortunately, we understand the microscopic structure of fluids much
better than quantum gravity.
For many fluids, the first deviations from the Euler equation at short 
distances manifest themselves in a change of the dispersion relation
for sound. 
In the case of Bose-Einstein condensates, for example, the dispersion
relation reads (for a fluid at rest)
\bea
\label{dispersion}
\omega^2=c_{\rm s}^2k^2+\frac{\hbar^2k^4}{4m^2}
\,,
\ea
where $m$ is the mass of the condensed particles \cite{Dalfovo}. 
(Note that this quantity does not appear at all in the Euler
equation.) 
For large wave-numbers $k$, the dispersion relation approaches the
non-relativistic free-particle limit $E=\hbar^2k^2/(2m)$ and thus 
becomes super-sonic, i.e., group and phase velocity exceed the speed of 
sound $v_{\rm group}=d\omega/dk>v_{\rm phase}=\omega/k>c_{\rm s}$. 
Hence, a wave-packet with such large wave-numbers is able to overcome
the frame-dragging with the flow velocity $v$ and thereby to approach
the sonic horizon from the inside, see Fig.~\ref{laval-BEC}. 
During that process, the inhomogeneity of $v$ (i.e., $v$ is smaller on 
the front end of the wave-packet than on its rear end) stretches the
wave-packet and reduces its wave-number
(analogous to the gravitational red-shift) and thereby lowers its
group velocity.
Eventually, the wave-packet gets ``ripped apart'' and one part
(the Hawking radiation) escapes into the exterior region whereas the
remaining part (the in-falling partner particle) is swept away into
the interior domain.
Assuming that the wave-packet was in its ground state initially
(for large $k$), the combined quantum state of the outgoing Hawking
quanta and their in-falling partner particles is a pure state -- but
the reduced density matrix of the Hawking radiation alone is thermal
%###
with the Hawking temperature, provided that the knee frequency, i.e.,
the frequency where the dispersion relation deviates from the linear
behaviour (the analogue of the Planck scale), is much bigger than the
Hawking temperature, see., e.g., \cite{Universality,Corley}. 

These considerations can be generalised to almost arbitrary dispersion
%###
relations (including the sub-sonic case\footnote{For a sub-sonic
  dispersion relation $d\omega/dk<c_{\rm s}$, the
  initial short-wavelength wave-packet originates from the exterior 
  region (r.h.s.~in Fig.~\ref{laval-BEC}) where $v<c_{\rm s}$.
  However, since its group velocity is too small, it is swept 
  towards the horizon and thereby stretched. Eventually, it also gets 
  ``ripped apart'' into two final long-wavelength wave-packets, which
  are basically the same as in the super-sonic case.}) 
and show that the Hawking effect survives in a large class of systems
%###
for which the dispersion relation $\omega(k)$ behaves
regularly\footnote{Such as Eq.~(\ref{dispersion}) or
  $\omega^2=c_{\rm s}^2k^2/(1+\eta^2k^2)$; but not
  $\omega^2=c_{\rm s}^2k^2/(1-\eta^2k^2)$, which becomes singular at 
  $k=1/\eta$ [manuscript in preparation].}  
for all $k$ 
%
%does not behave too violently at large $k$ and 
%
and becomes linear $\omega=c_{\rm s}k$ at the typical frequency of the 
outgoing Hawking radiation. 
However, so far everything referred to the local fluid frame -- which
is analogous to the local freely falling frame in gravity.
For example, the frequency $\omega$ in Eq.~(\ref{dispersion}) is
measured with respect to the co-moving fluid frame.
The (conserved) frequency $\Omega$ in the laboratory frame -- which
would correspond to the global rest frame of the black hole in gravity
-- is locally Doppler shifted $(\Omega+vk)^2=\omega^2(k)$.
Now, if we allow for some interaction between the fluid and the walls
of the nozzle (which are static in the laboratory frame), we would
include an effective friction term $\gamma$ into this dispersion
relation  
\bea
(\Omega+vk)^2+i\gamma\Omega=\omega^2(k)
\,.
\ea
Calculating the solutions of this quadratic equation, we see that the
frequency $\Omega$ acquires an imaginary part which is proportional to
$\gamma$ multiplied by the difference between the flow velocity 
$v_{\rm fluid}$ and the sound speed 
(or, more precisely, its phase velocity $v_{\rm phase}=\omega/k$).  
Therefore, for sub-sonic flow velocities, the interaction with the
wall would just induce a damping of the sound waves (as one would
expect). 
For super-sonic flow velocities, however, the effect reverses its
sign and thus the phonons are not damped but amplified!
This phenomenon is known as Miles instability \cite{Miles} and is
mainly responsible for the generation of water waves by
wind.\footnote{If the 
  wind blows faster than the phase velocity of the water waves, they
  are not damped but amplified by their friction with air. Since
  deep-water waves with large wavelengths are faster than those with
  shorter wavelengths, the velocity of the wind determines the maximum
  size of the generated waves.}
This amplification mechanism may occur in the left half of the
de~Laval nozzle in Fig.~\ref{laval-BEC}, where the flow velocity
exceeds the sound speed (analogous to the interior region of the black
hole). 
Consequently, even if the wave-packet in Fig.~\ref{laval-BEC} started 
out in its ground state (with respect to the fluid frame) for large
$k$, a coupling to the wall would generate excitations on its way
towards the horizon 
(due to $v_{\rm fluid} \approx v_{\rm group} > v_{\rm phase}$)
and these alterations of the initial quantum state would cause
deviations from Hawking's result.  
For example, if the interaction with the wall causes the transition of
the wave-packet from its ground state with respect to the fluid frame
towards its ground state with respect to the rest frame of the walls 
(let us assume fermionic quasi-particles for the moment in order to
make the Bogoliubov coefficients $|\alpha_k^2|+|\beta_k^2|=1$ and 
single-mode Hamiltonians bounded from above and below), there
would be no Hawking radiation at all (Boulware state). 

%%%%%%%%%%%%%%%%%%%%%%%%%%%%%%%%%%%%%%%%%%%%%%%%%%%%%%%%%%%%%%%%%%%%%%%%%%%%%%%
%%%%%%%%%%%%%%%%%%%%%%%%%%%%%%%%%%%%%%%%%%%%%%%%%%%%%%%%%%%%%%%%%%%%%%%%%%%%%%%
\section{Apparent horizon}
%%%%%%%%%%%%%%%%%%%%%%%%%%%%%%%%%%%%%%%%%%%%%%%%%%%%%%%%%%%%%%%%%%%%%%%%%%%%%%%
%%%%%%%%%%%%%%%%%%%%%%%%%%%%%%%%%%%%%%%%%%%%%%%%%%%%%%%%%%%%%%%%%%%%%%%%%%%%%%%

\begin{figure}[ht]
%\mbox{
\epsfxsize=6cm\epsffile{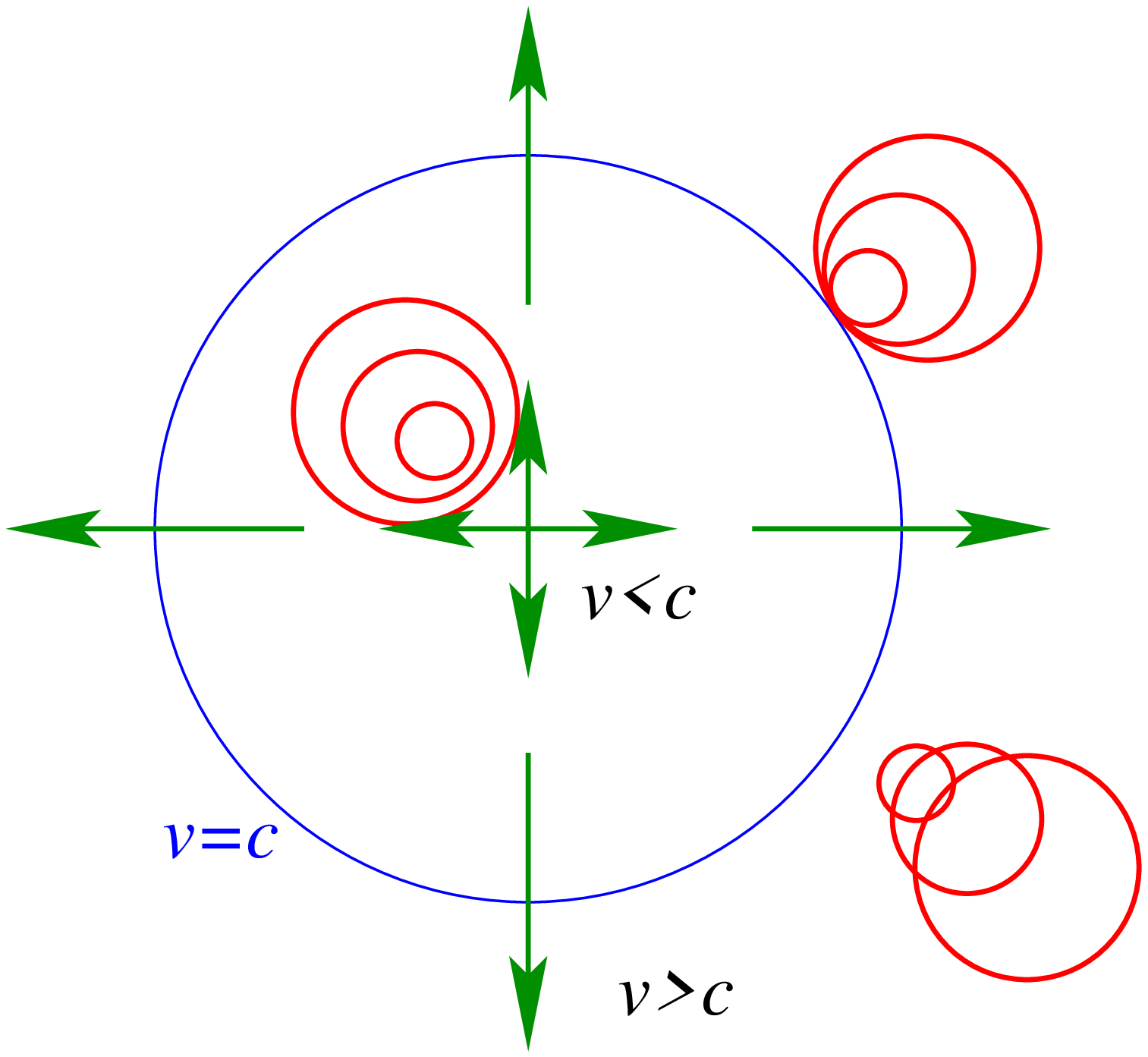}
\epsfxsize=6cm\epsffile{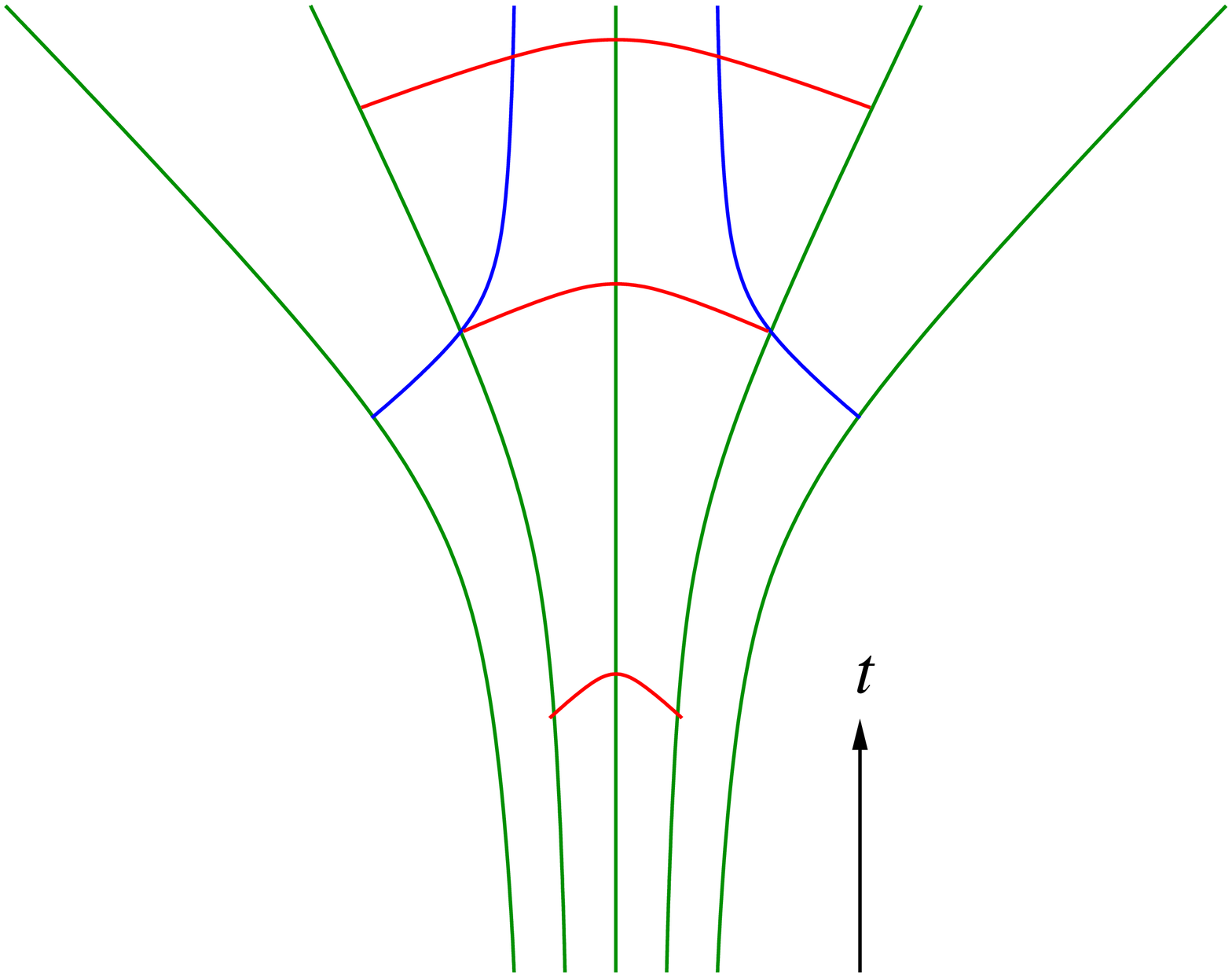}
%}
\caption{Sketch of an instantaneous snap-shot (top) and space-time
  diagram (bottom) of an expanding fluid. 
  The green lines (bottom) denote fluid particle trajectories and the
  green arrows (top) the associated flow velocities (measured with
  respect to the centre of the fluid, which is at rest). The apparent
  horizon (blue) occurs where these flow velocities exceed the speed of
  sound. Obviously, no sound wave (red circles) can reach the centre
  from the region outside (at that instant of time). The three stages
  of the evolution of the sound modes (red lines) are sketched from
  bottom to top in the lower picture: Initially, they lie far within
  the horizon and hence oscillate almost freely. Later, they are
  stretched by the expansion of the fluid and cross the
  horizon. Afterwards, crest and trough	of the wave are separated by
  the horizon and hence cannot exchange energy anymore, i.e., the
  oscillation freezes. As a result, the quantum state of this phonon
  mode cannot adapt to the expansion anymore and behaves
  non-adiabatically (amplification of quantum fluctuations by
  squeezing). 
}
\label{instantan}
\end{figure}

Since event horizons are coordinate and observer independent, they can
only occur in inhomogeneous systems.
In cosmology, on the other hand, other horizon concepts are more
suitable for describing the loss of causal connection within an
expanding universe.  
Recalling the analogy described in Section~\ref{analogy}, it turns out
that an expanding universe can be modelled by a homogeneously and
isotropically expanding fluid, see Fig.~\ref{instantan}. 
In this situation, we may specify the effective metric in
Eq.~(\ref{PGL})  
\begin{eqnarray}
\label{expanding}
ds_{\rm eff}^2
=
\frac{\varrho_0}{c_{\rm s}}
\left(
[c_s^2 - {\vau}_0^2]\,dt^2
+ 2 {\vau}_0 \cdot d{\f{r}}\,dt - d{\f{r}}^2
\right)
\,,
\end{eqnarray}
via inserting $\varrho_0(t,\f{r})\to\varrho_0(t)$ and hence 
$c_{\rm s}(t,\f{r})\to c_{\rm s}(t)$ [homogeneity] 
as well as $\vau(t,\f{r})\to a(t)\f{r}$ [isotropy]. 
Diagonalising the above metric by the transformation to co-moving
coordinates via ${\f{r}}=b(t){\f{R}}$, where $a=\dot b/b$, 
we arrive at the usual Friedmann-Robertson-Walker metric 
\begin{eqnarray}
\label{co-moving}
ds_{\rm eff}^2
=
\frac{\varrho_0(t)}{c_{\rm s}(t)}
\left(
c_s^2(t)\,dt^2 - b^2(t)\,d{\f{R}}^2
\right)
\,.
\end{eqnarray}
For a static fluid, every point may send/receive sound waves to/from
any other point. 
However, if the fluid is expanding such that its flow velocity exceeds
the speed of sound at some point, this is obviously no longer true,
see Fig.~\ref{instantan}.  

Employing the analogy to gravity, this border-line corresponds to an 
apparent horizon (i.e., the outermost trapped surface), which denotes
the momentary loss of causal connection. 
In view of the quantitative correspondence described in
Section~\ref{analogy}, the initial quantum fluctuations of the phonon
modes are frozen and amplified by the background expansion -- in
complete analogy to gravity.
In our standard model of cosmology, precisely this mechanism
(oscillation $\to$ horizon crossing $\to$ freezing $\to$ squeezing) is
responsible for the generation of the seeds for structure formation
out of the initial quantum fluctuations of the inflaton field during
inflation.  
Traces of these frozen fluctuations can still be observed today in the
anisotropies (on the level of $10^{-5}$) of the cosmic microwave
background radiation. 
As demonstrated above, the same effect occurs in expanding fluids such
as Bose-Einstein condensates, see, e.g., \cite{Expanding}. 
In this case, the amplified quantum fluctuations result in small
inhomogeneities, which are on the percent-level for realistic
parameters (i.e., the generated contrast is even bigger than in the
cosmological case) and could become measurable in near-future
experiments.  
%###
Besides the experimental point of view, one may use this analogy to
study the impact of the dispersion relation etc. 

%%%%%%%%%%%%%%%%%%%%%%%%%%%%%%%%%%%%%%%%%%%%%%%%%%%%%%%%%%%%%%%%%%%%%%%%%%%%%%%
%%%%%%%%%%%%%%%%%%%%%%%%%%%%%%%%%%%%%%%%%%%%%%%%%%%%%%%%%%%%%%%%%%%%%%%%%%%%%%%
\section{Particle horizon}
%%%%%%%%%%%%%%%%%%%%%%%%%%%%%%%%%%%%%%%%%%%%%%%%%%%%%%%%%%%%%%%%%%%%%%%%%%%%%%%
%%%%%%%%%%%%%%%%%%%%%%%%%%%%%%%%%%%%%%%%%%%%%%%%%%%%%%%%%%%%%%%%%%%%%%%%%%%%%%%

\begin{figure}[ht]
%\mbox{
%\epsfxsize=7cm\epsffile{qpt-dyn.eps}
%\hspace{1cm}
\epsfxsize=5cm\epsffile{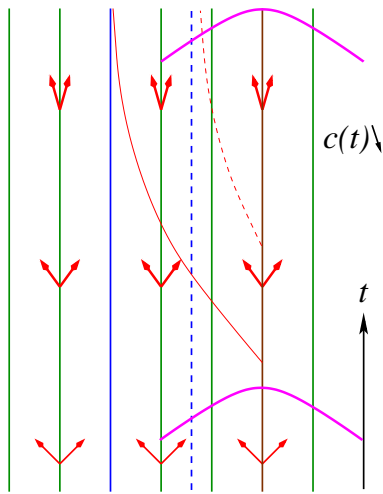}
%}
\label{particle}
\caption{Sketch of the 
%level structure for a phase transition at zero temperature (left) and 
space-time diagram of the sweep through such a phase transition and the 
emergence of a particle horizon. 
%
%In the left picture, the
%  energy $E$ is plotted as a function of some (time-dependent)
%  external parameter $g(t)$, such as pressure or magnetic field. The
%  green lines denote two competing vacuum states $\ket{\Psi_>}$ and
%  $\ket{\Psi_<}$ (e.g., para- and ferro-magnetic phase) and their
%  intersection denotes the critical point $g=g_c$. Some of the
%  quasi-particles (red dotted lines) may be unaffected by the
%  transition ($\chi$) but others ($\phi$) may become unstable after
%  crossing the critical point. Since their energy gaps vanish and
%  response times diverge at $g=g_c$, the actual quantum state
%  $\ket{\Psi(t)}$ (brown dashed line) behaves non-adiabatically
%  [cf.~Eq.~(\ref{adiabatic})] and deviates from the real ground state
%  when approaching the transition via a time-dependent $g(t)$. 
%
%The emergence of a particle horizon is sketched in the right
%  picture. 
%
Again, the green lines denote fluid particle trajectories
  and the red arrows are the time-dependent sound cones. If the speed
  of sound decreases fast enough, a sound wave (red curve) can only
  travel a finite distance and hence the region beyond the particle
  horizon (blue line) is causally disconnected from the brown
  trajectory. Since the size of the particle horizon shrinks
  constantly (see dashed red curve and dashed blue line for a later
  time), every sound mode (magenta curve) crosses the horizon at some
  instant and freezes.  
}
\end{figure}

Recalling the effective metric in Eq.~(\ref{PGL}), we see that an
expanding/contracting universe can also be modelled by a fluid at rest
$\vau_0=0$ -- provided that the speed of sound changes with time
\begin{eqnarray}
\label{rest}
ds_{\rm eff}^2
=
\varrho_0c_{\rm s}(t)\,dt^2
-
\frac{\varrho_0}{c_{\rm s}(t)}
\,d{\f{r}}^2
\,. 
\end{eqnarray}
In this case, an apparent horizon (which depends on the chosen time 
slicing) cannot be defined with respect to the laboratory time $t$
and, therefore, yet another horizon concept is more useful.
If the speed of sound decreases fast enough, sound waves can only
travel a finite distance, see Fig.~\ref{particle}
\bea
\label{p-hor}
r_{\rm horizon}(t)=\int\limits_t^\infty dt'\,c_{\rm s}(t') 
\,,
\ea
and a particle horizon arises.  
Since the effective metrics in Eqs.~(\ref{expanding}) and (\ref{rest})
are related to each other 
[for suitable dynamics $b(t)$ and $c_{\rm s}(t)$] via a simple
coordinate transformation, they describe the same physical effects --
even though the underlying condensed-matter systems are very
different (expanding fluid vs.\ fluid at rest). 
This observation demonstrates the universality of the geometrical
concepts and allows us to predict the same amplification mechanism
(oscillation $\to$ horizon crossing $\to$ freezing $\to$ squeezing) 
as in the previous Section.
For a fluid at rest, the wavelength of a given mode remains constant,
but the size of the particle horizon shrinks constantly [as can be
seen from Eq.~(\ref{p-hor})] and thus every mode crosses the horizon
at some time, see Fig.~\ref{particle}. 

A prototypical example for a condensed-matter system in which the
speed of sound (or other quasi-particles) goes down to zero is a sweep
through a zero-temperature phase transition by means of a
time-dependent external parameter $g(t)$, see Fig.~\ref{particle}.  
With the same assumptions as in Eq.~(\ref{general}), the effective
Hamiltonian density for the quasi-particle excitations $\phi$ within
the homogeneous and isotropic sample can be cast into the general form 
\bea
{\mathcal H}=\frac12\left(\alpha\Pi^2+\beta(\na\phi)^2\right)
\,,
\ea
where $\Pi$ is the canonically conjugate momentum density and 
$\alpha$ and $\beta$ are external parameters which depend on $g(t)$
and thereby on time.
If the quasi-particle excitations $\phi$ are stable before the
transition $g<g_c$ (i.e., $\alpha,\beta>0$) but become unstable after
crossing the critical point $g>g_c$ (i.e., they trigger the decay of
the false vacuum down to the real ground state), at least one of the
two parameters must change its sign at $g=g_c$.  
Consequently, the speed of sound $c^2_{\rm s}=\alpha\beta$ goes to
zero at the transition (and becomes imaginary afterwards -- indicating
the aforementioned instability), i.e., the $\phi$-modes become
arbitrarily soft when approaching the transition. 
The frozen and amplified quantum fluctuations (as long as they are in
the linear low-energy regime) can then be calculated in complete
analogy to cosmology \cite{qpt,phase}. 
The other way around, one might use these condensed-matter analogues
as toy models for quantum gravity in order to address questions such
as the quantum back-reaction problem or possible low-energy signatures
of Planck-scale physics, see, e.g., \cite{Queisser}. 

%%%%%%%%%%%%%%%%%%%%%%%%%%%%%%%%%%%%%%%%%%%%%%%%%%%%%%%%%%%%%%%%%%%%%%%%%%%%%%%
%%%%%%%%%%%%%%%%%%%%%%%%%%%%%%%%%%%%%%%%%%%%%%%%%%%%%%%%%%%%%%%%%%%%%%%%%%%%%%%
\section{Lessons for quantum gravity?}
%%%%%%%%%%%%%%%%%%%%%%%%%%%%%%%%%%%%%%%%%%%%%%%%%%%%%%%%%%%%%%%%%%%%%%%%%%%%%%%
%%%%%%%%%%%%%%%%%%%%%%%%%%%%%%%%%%%%%%%%%%%%%%%%%%%%%%%%%%%%%%%%%%%%%%%%%%%%%%%

What are the lessons for quantum gravity to be learned from these
analogies? 
We found that the Hawking effect -- even though its original
derivation is based on an over-extrapolation of the semi-classical
analysis -- seems to be fairly robust against alterations of the
physics at ultra-high (Planckian) energies for a large class of
models. 
On the other hand, there are also physically reasonable examples which
display strong deviations from Hawking's prediction.
Pictorially speaking, it appears to be important whether the
``space-time foam'' of quantum gravity is freely falling into the
black hole (as everything else) or whether it is fixed with respect
to the global rest frame of the black hole and just ``tells'' the
matter to fall down.
In the first case, one would expect\footnote{But then one should worry
  what happens with the ``space-time foam'' at the singularity --
  which might be related to the mode-generation problem in cosmology
  and the black-hole information paradox.} 
Hawking's prediction to be valid 
%###
(if the microscopic behaviour is not too violent, i.e., adiabatic);
whereas, in the second situation, one could easily imagine deviations
from Hawking's result.  
\\
Furthermore, we observed that the Hawking effect is fairly independent
of the Einstein equations and just requires the existence of an
emergent (event) horizon. 
The concept of black-hole entropy, on the other hand, cannot be
applied to the sonic analogues and probably relies on some features of
the Einstein equations. 
In the light of this observation, the frequently suggested
interpretation of the (non-universal) black-hole entropy via the
entanglement of the (universal) Hawking radiation appears a bit
dubious. 
\\
In contrast to effective metrics and horizons, which emerge quite
naturally in many condensed matter systems, it is very hard to
reproduce the principle of equivalence, which might give us some hints
about the underlying structure of quantum gravity.  
Speculating a bit further, one may ask whether {\em every} microscopic
model which incorporates general covariance in a suitable form would
lead to the Einstein equations as low-energy effective theory and
hence constitute a valid candidate for quantum gravity. 
(Demanding covariance on a microscopic level could perhaps explain why 
the direct search for Lorentz violations was negative so far.) 
As we know from condensed matter, many fluids with totally different
microscopic structures are described by the same Euler equation at
large distances and low energies -- the same could perhaps be true for
the Einstein equations. 

%%%%%%%%%%%%%%%%%%%%%%%%%%%%%%%%%%%%%%%%%%%%%%%%%%%%%%%%%%%%%%%%%%%%%%%%%%%%%%%
\section*{Acknowledgments}
%%%%%%%%%%%%%%%%%%%%%%%%%%%%%%%%%%%%%%%%%%%%%%%%%%%%%%%%%%%%%%%%%%%%%%%%%%%%%%%

R.~S.~acknowledges valuable discussions with Bill Unruh and many
others 
as well as financial support by the Emmy-Noether Programme 
of the German Research Foundation (DFG, SCHU~1557/1-2,3).

%$^*$\,{\sf schuetz@theory.phy.tu-dresden.de}

%%%%%%%%%%%%%%%%%%%%%%%%%%%%%%%%%%%%%%%%%%%%%%%%%%%%%%%%%%%%%%%%%%%%%%%%%%%%%%%
\addcontentsline{toc}{section}{References}
%%%%%%%%%%%%%%%%%%%%%%%%%%%%%%%%%%%%%%%%%%%%%%%%%%%%%%%%%%%%%%%%%%%%%%%%%%%%%%%

%%%%%%%%%%%%%%%%%%%%%%%%%%%%%%%%%%%%%%%%%%%%%%%%%%%%%%%%%%%%%%%%%%%%%%%%%%%%%%%
%%%%%%%%%%%%%%%%%%%%%%%%%%%%%%%%%%%%%%%%%%%%%%%%%%%%%%%%%%%%%%%%%%%%%%%%%%%%%%%
%%%%%%%%%%%%%%%%%%%%%%%%%%%%%%%%%%%%%%%%%%%%%%%%%%%%%%%%%%%%%%%%%%%%%%%%%%%%%%%
%%%%%%%%%%%%%%%%%%%%%%%%%%%%%%%%%%%%%%%%%%%%%%%%%%%%%%%%%%%%%%%%%%%%%%%%%%%%%%%
%%%%%%%%%%%%%%%%%%%%%%%%%%%%%%%%%%%%%%%%%%%%%%%%%%%%%%%%%%%%%%%%%%%%%%%%%%%%%%%
%%%%%%%%%%%%%%%%%%%%%%%%%%%%%%%%%%%%%%%%%%%%%%%%%%%%%%%%%%%%%%%%%%%%%%%%%%%%%%%
%%%%%%%%%%%%%%%%%%%%%%%%%%%%%%%%%%%%%%%%%%%%%%%%%%%%%%%%%%%%%%%%%%%%%%%%%%%%%%%
%%%%%%%%%%%%%%%%%%%%%%%%%%%%%%%%%%%%%%%%%%%%%%%%%%%%%%%%%%%%%%%%%%%%%%%%%%%%%%%
%%%%%%%%%%%%%%%%%%%%%%%%%%%%%%%%%%%%%%%%%%%%%%%%%%%%%%%%%%%%%%%%%%%%%%%%%%%%%%%
%%%%%%%%%%%%%%%%%%%%%%%%%%%%%%%%%%%%%%%%%%%%%%%%%%%%%%%%%%%%%%%%%%%%%%%%%%%%%%%
%%%%%%%%%%%%%%%%%%%%%%%%%%%%%%%%%%%%%%%%%%%%%%%%%%%%%%%%%%%%%%%%%%%%%%%%%%%%%%%
%%%%%%%%%%%%%%%%%%%%%%%%%%%%%%%%%%%%%%%%%%%%%%%%%%%%%%%%%%%%%%%%%%%%%%%%%%%%%%%
%%%%%%%%%%%%%%%%%%%%%%%%%%%%%%%%%%%%%%%%%%%%%%%%%%%%%%%%%%%%%%%%%%%%%%%%%%%%%%%
%%%%%%%%%%%%%%%%%%%%%%%%%%%%%%%%%%%%%%%%%%%%%%%%%%%%%%%%%%%%%%%%%%%%%%%%%%%%%%%
%%%%%%%%%%%%%%%%%%%%%%%%%%%%%%%%%%%%%%%%%%%%%%%%%%%%%%%%%%%%%%%%%%%%%%%%%%%%%%%
%%%%%%%%%%%%%%%%%%%%%%%%%%%%%%%%%%%%%%%%%%%%%%%%%%%%%%%%%%%%%%%%%%%%%%%%%%%%%%%
\end{document}